\documentclass[11pt]{article}
\usepackage{graphicx,amsmath,amsfonts,amssymb,dcolumn,times}

\begin{document}

\title{\textbf{Formal aspects of spin currents in material media}}
\author{\textbf{T.~R.~S.~Santos$^1$}\thanks{tiagoribeiro@if.uff.br}\ ,\ \textbf{R.~F.~Sobreiro$^1$}\thanks{sobreiro@if.uff.br}\ ,\ \textbf{V.~J.~Vasquez Otoya$^2$}\thanks{victor.vasquez@ifsudestemg.edu.br}\\\\
\textit{{\small $^1$UFF $-$ Universidade Federal Fluminense,}}\\
\textit{{\small Instituto de F\'{\i}sica, Campus da Praia Vermelha,}}\\
\textit{{\small Avenida General Milton Tavares de Souza s/n, 24210-346,}}\\
\textit{{\small Niter\'oi, RJ, Brasil.}}\and
\textit{{\small $^2$IFSEMG $-$ Instituto Federal de Educa\c{c}\~ao, Ci\^encia e Tecnologia do Sudeste de Minas Gerais,}} \\
\textit{{\small Rua Bernardo Mascarenhas 1283, 36080-001,}}\\
\textit{{\small Juiz de Fora, MG, Brasil}}}
\date{}
\maketitle

\begin{abstract}
The relativistic generalization of the broken continuity equation that describes spin currents is exploited in general material media. The field theoretical techniques for the usual $U(1)$ gauge theory with fermions in material media is employed. The spin current equations are obtained from the study of the symmetries of electrodynamics in material media. Moreover, we obtain the conditions that constrain the spin current to be conserved. One of the novel results is that the analogue spin current for the photon can be also defined. In the latter case, the conditions are useful constraints on the fields and the properties of the media.
\end{abstract}

\section{Introduction}

Spintronics explores the spin properties of electrons independently of the charge transport \cite{Wolf,Awschalom,Zutic}. The main physical object that describes the transport of spin is the so called spin current. The understanding of its behavior are of major relevance for the development of new technologies. Nevertheless, the non-conservation of spin current continues to be a problem to be solved \cite{Qing,Vernes:2007,Sobreiro:2011ie}. The purpose of this work is to understand the fundamental nature of this non-conservation and to provide the conditions for the spin current to be conserved.

In an enlightening work by A.~Vernes, L.~Gy\"orffy and P.~Weinberger \cite{Vernes:2007}, the broken continuity equation for spin current was obtained from the non-relativistic limit of the time evolution of the Bargmann-Wigner operator \cite{Bargmann:1948ck,Itzykson:1980rh} within the Dirac Hamiltonian, resulting in
\begin{equation}
\frac{d\overrightarrow{s}}{dt}+\partial_i\overrightarrow{j}_i=\frac{e}{m}\overrightarrow{s}\times\overrightarrow{B}\;.\label{eq1}
\end{equation}
In expression \eqref{eq1}, $\overrightarrow{s}=\phi^\dagger\overrightarrow{\sigma}\phi$ is the spin density, $\overrightarrow{j}_i$ is the spin current, $\phi$ is the non-relativistic electron wave function and the \emph{rhs} is the usual microscopic Landau-Lifshitz torque. Equation \eqref{eq1} can be derived from the study of the time evolution of $\overrightarrow{s}$ within Pauli equation for the electron \cite{Stiles}. Escaping from standard electromagnetism, equation \eqref{eq1} can also be obtained from the flavor current of weak interactions, \emph{i.e.}, the $SU(2)$ gauge current \cite{Dartora:2008ccc, Dartora:2010zz}. Thus, spin current flow can actually be casted into a continuity equation, even though standard electromagnetism based on the $U(1)$ gauge symmetry is substituted by its next non-Abelian extension. It is fair to mention that Equation \eqref{eq1} first appeared in \cite{Sokolov:1986nk}, although no reference to spin currents has been made. Finally, In \cite{Sobreiro:2011ie}, a formal study of the origin of Equation \eqref{eq1} was performed by employing field theory techniques. In fact, the $U(1)$ gauge theory for electromagnetism in vacuum \cite{Itzykson:1980rh,Barut} was considered. In particular, the relativistic generalization of Equation \eqref{eq1} was obtained from the combination of the two most important symmetries of electromagnetism, namely, Lorentz and gauge symmetries. The relevant sector of the Lorentz symmetry is the restricted subgroup $L(1,3)\subset SO(1,3)$ called little group whose generators can be combined into a Casimir operator which is associated to spin eigenvalues. The same technique is applied here for the electromagnetism in a generic material media. Moreover, a spin-current analogue for the photon \cite{Sobreiro:2011ie} is also defined and discussed.

The starting point of this work is the Minkowski-Maxwell-Dirac action \cite{Barut,Post}. This action describes electrodynamics in a general material medium and electrons coupled to the electromagnetic field. The currents associated with gauge symmetry, chiral asymmetry, and little group symmetry are easily obtained. The latter is commonly known as Bargmann-Wigner current whose generators are precisely those associated with the values of spin in the same way that the generators of the subgroup of translations are associated to the mass value of particles. Although conserved, the Bargmann-Wigner current is not gauge invariant and then, it is hard to be associated with physical observables. Thus, by evoking the gauge principle for electrodynamics, we restore the gauge invariance of the Bargmann-Wigner. The price that is payed is that the spin current is no longer conserved. In that approach, the electromagnetic field is dynamical and a similar equation for electromagnetic fields is obtained. On the other hand, the broken continuity equation for the electronic sector is the same as the vacuum case \cite{Sobreiro:2011ie} while the photonic equation changes due to the properties of the medium. Finally, we explore the conditions for those currents to be conserved.

The letter is organized as follows: In Sect.~2 we study the starting action, its relevant symmetries and the respective conserved currents. In Sect.~3 we construct the gauge invariant spin currents and obtain their respective broken continuity equations. The space and time decomposition of these currents and the conservation conditions are obtained in Sect.~4. Finally, our conclusions and a discussion are displayed in Sect.~5.

\section{Electrodynamics in general material media}

We start with the usual action for electrodynamics in material media \cite{Post}
\begin{equation}
S=\int{d^4x}\;\overline{\psi}\left(i\hbar c\gamma^\mu D_\mu-mc^2\right)\psi-\frac{1}{4}\int{d^4x}\;G^{\mu\nu}F_{\mu\nu}\;,\label{action}
\end{equation}
where the field $\psi$ is a spinor field describing electron excitations and $\overline{\psi}$ its adjoint, $\overline{\psi}=\psi^\dagger\gamma^0$. The Clifford algebra $\left\{\gamma^\mu,\gamma^\nu\right\}=2\eta^{\mu\nu}$ allows the usage of Dirac representation for the $\gamma$-matrices and the metric tensor is defined with negative signature, $\eta=\mathrm{diag}(+1,-1,-1,-1)$. Useful extra quantities are $\gamma^5=\gamma_5=i\gamma^0\gamma^1\gamma^2\gamma^3$ and $\sigma^{\mu\nu}=\frac{i}{2}\left[\gamma^\mu,\gamma^\nu\right]\label{sigma}$. The derivative $D_\mu=\partial_\mu+i\frac{e}{\hbar c}A_\mu$ is the covariant derivative and the field strength is defined as $F_{\mu\nu}=\partial_\mu A_\nu-\partial_\nu A_\mu$, where $A_\mu$ is the electromagnetic potential. The tensor $G^{\mu\nu}$ is an auxiliary antisymmetric tensor,
\begin{eqnarray}
G^{\mu\nu}=\frac{1}{2}\chi^{\mu\nu\alpha\beta}F_{\alpha\beta}\;,\label{g}
\end{eqnarray}
where $\chi^{\mu\nu\alpha\beta}$ is the constitutive pseudo-tensor whose symmetry properties are $\chi^{\mu\nu\alpha\beta}=-\chi^{\nu\mu\alpha\beta}=-\chi^{\mu\nu\beta\alpha}=\chi^{\alpha\beta\mu\nu}$. The inverse relation of \eqref{g} is
\begin{eqnarray}
F_{\mu\nu}=\frac{1}{2}\overline{\chi}_{\mu\nu\alpha\beta}G^{\alpha\beta}\;,\label{f}
\end{eqnarray}
where $\overline{\chi}_{\mu\nu\alpha\beta}$ is determined by $\overline{\chi}_{\mu\nu\alpha\beta}{\chi}^{\alpha\beta\gamma\delta}=2(\delta^{\gamma}_\mu\delta^\delta_\nu-\delta^{\gamma}_\nu\delta^\delta_\mu)$.

The fermionic field equations obtained from \eqref{action} are
\begin{eqnarray}
\left(i\gamma^\mu D_\mu-\frac{mc}{\hbar}\right)\psi&=&0\;,\nonumber\\
\overline{\psi}\left(i\gamma^\mu\overleftarrow{D}_\mu^\dagger+\frac{mc}{\hbar}\right)&=&0\;.\label{eqf}
\end{eqnarray}
For the electromagnetic field, the equations are
\begin{equation}
\partial_\nu G^{\nu\mu}=j^\mu_f\;,\label{eqb}
\end{equation}
where $j_f^\mu=e\overline{\psi}\gamma^\mu\psi$ is the fermionic charge current (see next Section). Equations \eqref{eqb} are recognized as the inhomogeneous Maxwell equations. The homogeneous Maxwell equations, $\partial_\nu
\widetilde{F}^{\nu\mu}=0$, are obtained from the topological properties of the theory, where the dual field strength is defined as $\widetilde{F}^{\mu\nu}=\frac{1}{2}\epsilon^{\mu\nu\alpha\beta}F_{\alpha\beta}$.

Obviously, the field $G^{\mu\nu}$ is composed by the fields $\vec{D}$ and $\vec{H}$ while the field strength $F^{\mu\nu}$ has $\vec{E}$ and $\vec{B}$ as components. Thus,
\begin{equation}
G^{\mu\nu}\equiv\begin{pmatrix}
0&cD^1&cD^2&cD^3\\
-cD^1&0&-H^3&H^2\\
-cD^2&H^3&0&-H^1\\
-cD^3&-H^2&H^1&0
\end{pmatrix}\;,\;\;\;
F^{\mu\nu}\equiv\begin{pmatrix}
0&-E^1/c&-E^2/c&-E^3/c\\
E^1/c&0&-B^3&B^2\\
E^2/c&B^3&0&-B^1\\
E^3/c&-B^2&B^1&0
\end{pmatrix}\;.\label{fs}
\end{equation}
The relation \eqref{g} can be \emph{unwrapped} as \cite{Post}
\begin{equation}
\begin{pmatrix}
\vec{D}\\
\vec{H}
\end{pmatrix}=\begin{pmatrix}
-\epsilon& \gamma\\
\gamma^\dagger& \zeta
\end{pmatrix}\begin{pmatrix}
-\vec{E}\\
\vec{B}
\end{pmatrix}\;,\label{dh}
\end{equation}
where the 3-dimensional tensors $\epsilon$, $\gamma$ and $\zeta=\mu^{-1}$ are related to electric permittivity, natural optical activity and magnetic permeability. In fact, from \eqref{g}, \eqref{fs} and \eqref{dh} we find
\begin{eqnarray}
D^i&=&\epsilon^{ik}E^k+\gamma^{ik}B^k\;,\nonumber\\
H^i&=&-{\gamma^*}^{ki}E^k+\zeta^{ik}B^k\;,
\end{eqnarray}
where
\begin{eqnarray}
\epsilon^{ik}&=&-\frac{1}{c^2}\chi^{0i0k}\;,\nonumber\\
\gamma^{ik}&=&\frac{1}{2c}\chi^{0ijl}\epsilon^{jlk}\;,\nonumber\\
{\gamma^*}^{ki}&=&\frac{1}{2c}\epsilon^{ijm}\chi^{jm0k}\;,\nonumber\\
\zeta^{ik}&=&\frac{1}{4}\epsilon^{ijl}\chi^{jlmn}\epsilon^{mnk}\;.
\end{eqnarray}

The action \eqref{action} displays a remarkable set of symmetries. Three of them is of great relevance in this work. We now discuss these symmetries. The first one is the $U(1)$ local symmetry which is characterized by the gauge transformations
\begin{eqnarray}
\delta_g\psi&=&-i\frac{e}{\hbar c}\alpha\psi\;,\nonumber\\
\delta_g\overline{\psi}&=&i\frac{e}{\hbar c}\alpha\overline{\psi}\;,\nonumber\\
\delta_gA_\mu&=&\partial_\mu\alpha\;,\label{gt}
\end{eqnarray}
where $\alpha$ is a spacetime dependent parameter. The action \eqref{action} is invariant under gauge transformations and the associated conserved current is $j_f^\mu=e\overline{\psi}\gamma^\mu\psi$, which expresses the conservation of the electric charge. Another well known fact in electrodynamics is the presence of the chiral symmetry for massless fermions. The chiral transformations
are defined as $\delta_c\psi=-i\frac{\alpha}{\hbar c}\gamma^5\psi$, $\delta_c\overline{\psi}=-i\frac{\alpha}{\hbar c}\overline{\psi}\gamma^5$ and $\delta_cA_\mu=0$, where $\alpha$ is now a constant parameter. The chiral non-conserved current is easily computed as
\begin{equation}
S^\mu=\overline{\psi}\gamma^\mu\gamma^5\psi\;,\label{chiralb}
\end{equation}
which leads to the broken continuity equation $\partial_\mu S^\mu=2i\frac{mc}{\hbar}\overline{\psi}\gamma^5\psi$.

The main set of transformations relevant to this work originates from the Poincar\'e group, $ISO(1,3)=SO(1,3)\ltimes\mathbb{R}^4$. The generators of the Poincar\'e group are denoted by $P_\mu=i\hbar\partial_\mu$ for translations and $J_{\mu\nu}=L_{\mu\nu}+I_{\mu\nu}$ for the Lorentz sector. Here, $L_{\mu\nu}$ is taken as the angular momentum part, $L_{\mu\nu}=i\hbar\left(x_\mu\partial_\nu-x_\nu\partial_\mu\right)/2$, while $I_{\mu\nu}$ is associated with the internal angular momentum. The Poincar\'e algebra can be described by:
\begin{eqnarray}
\left[P_\mu,P_\nu\right]&=&0\;,\nonumber\\
\left[J_{\mu\nu},J_{\alpha\beta}\right]&=&-\frac{i\hbar}{2}\left(\eta_{\mu\alpha}J_{\nu\beta}-\eta_{\mu\beta}J_{\nu\alpha}-\eta_{\nu\alpha}J_{\mu\beta}+ \eta_{\nu\beta}J_{\mu\alpha}\right)\;,\nonumber\\
\left[J_{\mu\nu},P_\alpha\right]&=&\frac{i\hbar}{2}\left(\eta_{\alpha\nu}P_\mu-\eta_{\alpha\mu}P_\nu\right)\;.\label{sl2c}
\end{eqnarray}
The so called little group, $L(1,3)\subset SO(1,3)$, can be understood as the set of Lorentz transformations that keeps invariant the linear momentum. This subsector of the Poincar\'e group is described by the generator
\begin{equation}
W^\mu=-\frac{1}{2\hbar}\epsilon^{\mu\nu\alpha\beta}J_{\nu\alpha}P_\beta=-\frac{1}{2\hbar}\epsilon^{\mu\nu\alpha\beta}I_{\nu\alpha}P_\beta\;,
\end{equation}
which is the Pauli-Lubanski vector. The generator $W^\mu$ has the following properties:
\begin{eqnarray}
\left[W^\mu,W^\nu\right]&=&\epsilon^{\mu\nu\alpha\beta}P_\alpha W_\beta\;,\nonumber\\
\left[J_{\mu\nu},W^\alpha\right]&=&-\frac{i\hbar}{2}\left(\delta^\alpha_\mu W_\nu-\delta^\alpha_\nu W_\mu\right)\;,\nonumber\\
\left[W^\mu,P_\alpha\right]&=&0\;,\label{little}
\end{eqnarray}
emphasizing the subgroup character of the little group as well as the fact that $L(1,3)$ is a stability subgroup with respect to the Poincar\'e group.

For fermions, it is easy to find \cite{Itzykson:1980rh}
$I_{\mu\nu}=\hbar\sigma_{\mu\nu}/2$, providing
\begin{equation}
W^\mu_f=-\frac{1}{4}\epsilon^{\mu\nu\alpha\beta}\sigma_{\nu\alpha}P_\beta=\frac{i}{2}\gamma^5\sigma^{\mu\nu}P_\nu\;,\label{Pauli-Lubanski}
\end{equation}
where the index $f$ denotes its fermionic character. The little group transformations are
then
\begin{eqnarray}
\delta_l\psi&=&-i\frac{\omega_\mu}{\hbar}W^\mu_f\psi\;,\nonumber\\
\delta_l\overline{\psi}&=&-i\overline{\psi}\overleftarrow{W}^\mu_f\frac{\omega_\mu}{\hbar}\;,\label{little0}
\end{eqnarray}
with $\omega_\mu$ a set of constant real parameters. The related Noether current is a second rank tensor field, the so called Bargmann-Wigner tensor
\begin{equation}
T_f^{\mu\nu}=c\overline{\psi}\gamma^\mu W^\nu_f\psi\;.\label{bw0}
\end{equation}
Explicitly,
\begin{equation}
T_f^{\mu\nu}=-\frac{\hbar c}{2}\overline{\psi}\gamma^\mu\gamma^5\sigma^{\nu\alpha}\partial_\alpha\psi\;.\label{bw1}
\end{equation}

For the electromagnetic field, the Pauli-Lubanski vector reads\footnote{It follows from the spin part of the Lorentz generator for a vector field,
$\sigma^{\mu\nu\alpha\beta}=\hbar(\eta^{\mu\alpha}\eta^{\nu\beta}-\eta^{\mu\beta}\eta^{\nu\alpha})/2$.}
\begin{equation}
W_b^{\mu\nu\alpha}=-\frac{1}{2}\epsilon^{\mu\nu\alpha\beta}P_\beta\;,
\end{equation}
where the index $b$ characterizes its bosonic behavior. The little group transformation for $A_\mu$ is
\begin{equation}
\delta_lA_\mu=i\frac{\omega^\nu}{\hbar}W_{b\;\mu\nu\alpha}A^\alpha\;.\label{little1}
\end{equation}
Thus, for the vector field the corresponding Bargmann-Wigner current is
\begin{equation}
T_b^{\mu\nu}=\frac{1}{2}G^{\mu\alpha}\widetilde{F}_\alpha^{\phantom{\alpha}\nu}\;,\label{bw2}
\end{equation}
and the full Bargmann-Wigner conserved current is then
\begin{equation}
T^{\mu\nu}=T_f^{\mu\nu}+T_b^{\mu\nu}\;\;\bigg|\;\;\partial_\mu T^{\mu\nu}=0\;.\label{barg}
\end{equation}

\section{Gauge invariant currents}

It turns out that, in contrast to gauge and chiral currents, $T^{\mu\nu}$ is not a gauge invariant quantity, $\delta_gT_f^{\mu\nu}=-ie\overline{\psi}\gamma^\mu\left(W_f^\nu\alpha\right)\psi$. Thus, from the gauge principle, this current cannot be associated to a physical observable. Moreover, it is evident that it is only the fermionic sector $T^{\mu\nu}_f$ that breaks gauge symmetry. To circumvent this problem we generalize $T_f^{\mu\nu}$ to its simplest gauge invariant extension by replacing the ordinary derivative by the covariant one, \emph{i.e.}, the Pauli-Lubanski vector is replaced by
\begin{equation}
\mathcal{W}^\mu_f=-\frac{\hbar}{2}\gamma^5\sigma^{\mu\nu}D_\nu\;.
\end{equation}
The corresponding gauge invariant fermionic Bargmann-Wigner current is now
\begin{equation}
\mathcal{T}^{\mu\nu}_f=\frac{\hbar c}{2}\overline{\psi}\gamma^5\gamma^\mu\sigma^{\nu\alpha}D_\alpha\psi\;.\label{little1}
\end{equation}
Thus, since the electromagnetic sector is already gauge invariant, the full gauge invariant Bargmann-Wigner current is now gauge invariant, $\delta_g(\mathcal{T}_f^{\mu\nu}+T_b^{\mu\nu})=0$.

If in one hand we have the gauge invariance, in the other hand
$\mathcal{T}^{\mu\nu}$ is not conserved anymore. In fact, it can be shown that
\begin{equation}
\partial_\nu\mathcal{T}_f^{\nu\mu}=-\frac{e}{2}S_\nu
F^{\nu\mu}\;,\label{div1}
\end{equation}
where the field equations were used. For the bosonic sector it is easy to show that,
\begin{equation}
\partial_\nu T^{\nu\mu}_b=\frac{1}{2}j_{f\nu}\widetilde{F}^{\nu\mu}+\frac{1}{3}\widetilde{G}^{\mu\nu}\partial^\alpha F_{\alpha\nu}\;,\label{div2}
\end{equation}
where the field equations were used again. At equation \eqref{div2} we have used $\widetilde{G}^{\mu\nu}=\frac{1}{2}\epsilon^{\mu\nu\alpha\beta}G_{\alpha\beta}$.

Remarks: i.) equations \eqref{div1} and \eqref{div2} express the non-conservation of the gauge invariant
Bargmann-Wigner currents, and hold separately, since they are obtained independently of the continuity equation \eqref{barg}. ii.) The non-relativistic limit of \eqref{div1} reduces to the usual spin current equation \cite{Vernes:2007,Sobreiro:2011ie}.

One of the main problems with spin currents is the fact that they are not conserved quantities, \emph{i.e.}, in a generic system, equations \eqref{div1} and \eqref{div2} hold. The conditions for these currents to be conserved are then
\begin{equation}
S_\nu F^{\nu\mu}=0\;,\label{cond1}
\end{equation}
for the electron spin-current and
\begin{equation}
\frac{3}{2}j_{f\nu}\widetilde{F}^{\nu\mu}+\widetilde{G}^{\mu\nu}\partial^\alpha F_{\alpha\nu}=0\;,\label{cond2}
\end{equation}
for the bosonic spin-current. We will explore this issue in more detail at the next Section.

\section{Conserved currents}

To study the spin currents and their conservation it is convenient to decompose them into space and time sector. For that we define \cite{Sobreiro:2011ie}
\begin{eqnarray}
\mathcal{T}_{f}^{00}&=&-\frac{i\hbar c}{2}\psi^{\dag}\Sigma^{i}D_{i}\psi=-\frac{mc}{2}\mathcal{T}\;,\nonumber\\
\mathcal{T}_{f}^{i0}&=&-\frac{i\hbar c}{2}\psi^{\dag}\alpha^{i}\Sigma^{j}D_{j}\psi=-\frac{m}{2}\mathcal{T}^{i}\;,\nonumber\\
\mathcal{T}_{f}^{0i}&=&\frac{mc^2}{2}\psi^{\dag}\left(\beta\Sigma^{i}+\frac{i\hbar}{mc}\gamma^{5}D^{i}\right)\psi=\frac{mc}{2}\mathcal{J}^{i}\;,\nonumber\\
\mathcal{T}_{f}^{ij}&=&\frac{mc^2}{2}\psi^{\dag}\alpha^{i}\left(\beta\Sigma^{j}+\frac{i\hbar }{mc}\gamma^{5}D^{j}\right)\psi=\frac{m}{2}\mathcal{J}^{ij}\;,
\label{1p}
\end{eqnarray}
and
\begin{eqnarray}
T_b^{00}&=&\frac{c}{2}\vec{D}\cdot\vec{B}=c\mathcal{M}\;,\nonumber\\
T_b^{i0}&=&-\frac{1}{2}\left(\vec{H}\times\vec{B}\right)^i=\mathcal{M}^i\;,\nonumber\\
T_b^{0i}&=&-\frac{1}{2}\left(\vec{D}\times\vec{E}\right)^i=c\mathcal{N}^i\;,\nonumber\\
T_b^{ij}&=&-\frac{c}{2}D^iB^j+\frac{1}{2c}E^iH^j-\frac{1}{2c}E^kH^k\delta^{ij}=\mathcal{N}^{ij}\;.
\label{2p}
\end{eqnarray}
Thus, the non-conservation laws \eqref{div1} and \eqref{div2} decompose as
\begin{eqnarray}
\frac{\partial\mathcal{T}}{\partial t}+\vec{\nabla}\cdot\vec{\mathcal{T}}&=&-\frac{e}{mc}\vec{S}\cdot\vec{E}\;,\nonumber\\
\frac{\partial\vec{\mathcal{J}}}{\partial t}+\vec{\nabla}\cdot\stackrel{\leftrightarrow}{\mathcal{J}}&=&\frac{e}{m}\left(\frac{1}{c}S_0\vec{E}+\vec{S}\times\vec{B}\right)\;,\label{spin1}
\end{eqnarray}
and
\begin{eqnarray}
\frac{\partial\mathcal{M}}{\partial t}+\vec{\nabla}\cdot\vec{\mathcal{M}}&=&-\frac{1}{2}\vec{j}\cdot\vec{B}-\frac{1}{3}\vec{H}\cdot\left(\frac{1}{c^2}\frac{\partial\vec{E}}{\partial t}-\vec{\nabla}\times\vec{B}\right)\;,\nonumber\\
\frac{\partial\vec{\mathcal{N}}}{\partial t}+\vec{\nabla}\cdot\stackrel{\leftrightarrow}{\mathcal{N}}&=&-\frac{1}{2}\left(c\rho\vec{B}-\frac{1}{c}\vec{j}\times\vec{E}\right)+\frac{1}{3c}(\vec{\nabla}\cdot\vec{E})\vec{H}-\frac{c}{3}\vec{D}\times\left(\frac{1}{c^2}\frac{\partial\vec{E}}{\partial t}-\vec{\nabla}\times\vec{B}\right)\;,\label{spin2}
\end{eqnarray}
respectively. The second of \eqref{spin1} has as its non-relativistic limit the usual equation for spin currents where $\vec{\mathcal{J}}$ is the (relativistic) spin density and $\stackrel{\leftrightarrow}{\mathcal{J}}$ is the (relativistic) spin current \cite{Vernes:2007}. The same interpretation can be used for the second of \eqref{spin2} where $\vec{\mathcal{N}}$ is the (bosonic) density and $\stackrel{\leftrightarrow}{\mathcal{N}}$ is the bosonic current.

The conservation conditions \eqref{cond1} and \eqref{cond2} are reduced\footnote{The condition \eqref{cond3} is derived from the second of \eqref{spin1}. The first of \eqref{spin1} induces the condition $\vec{S}\cdot\vec{E}=0$, which follows naturally from \eqref{cond3}.} to
\begin{equation}
\vec{E}=-\frac{c}{S_0}\vec{S}\times\vec{B}\;.\label{cond3}
\end{equation}
and
\begin{eqnarray}
\frac{3}{2}\vec{j}\cdot\vec{B}+\vec{H}\cdot\left(\frac{1}{c^2}\frac{\partial\vec{E}}{\partial t}-\vec{\nabla}\times\vec{B}\right)&=&0\;,\nonumber\\
-\frac{3}{2}\left(c^2\rho\vec{B}-\vec{j}\times\vec{E}\right)+(\vec{\nabla}\cdot\vec{E})\vec{H}-c^2\vec{D}\times\left(\frac{1}{c^2}\frac{\partial\vec{E}}{\partial t}-\vec{\nabla}\times\vec{B}\right)&=&0\;,\label{cond4}
\end{eqnarray}
The condition \eqref{cond3} is a relatively simple requirement and accounts for the conservation of the electronic spin current. It imposes the conservation of the electronic spin current, independently of the media. Conditions \eqref{cond4}, on the other hand, are much more complicated and should account for the conservation of the photonic current. The second of \eqref{cond4} is actually the one that matters for the bosonic current conservation and it can be solved for $\vec{H}$,
\begin{equation}
\vec{H}=\frac{1}{(\vec{\nabla}\cdot\vec{E})}\left[\frac{3}{2}\left(c^2\rho\vec{B}-\vec{j}\times\vec{E}\right)+\vec{D}\times\left(\frac{\partial\vec{E}}{\partial t}-c^2\vec{\nabla}\times\vec{B}\right)\right]\;,\label{h}
\end{equation}
which is a quite complicated relation. Let us analyze simpler cases.

\subsection{Pure electronic case}

We consider the case which the electromagnetic field is external. In that case, $T_b^{\mu\nu}=0$ and equation \eqref{div1} is the one that describes the non-conservation of electronic spin current. It is clear that the condition for this current to be conserved is equation \eqref{cond1} whose decomposition is in expression \eqref{spin1}. The fact that \eqref{div1} is true for any kind of material media (including the vacuum) is a direct consequence of the action \eqref{action}. At this action, the spinor field directly couple with the electromagnetic field, \emph{i.e.}, the action \eqref{action} does not describe the interaction between the media and the electrons.

\subsection{Insulators}

Now, we consider a perfect insulator, \emph{i.e.}, there are no free electrons inside the material. Thus, $\mathcal{T}_f^{\mu\nu}=j^\mu_f=S^\mu=0$ and equation \eqref{div2} reduces to
\begin{equation}
\partial_\nu T^{\nu\mu}_b=\frac{1}{3}\widetilde{G}^{\mu\nu}\partial^\alpha F_{\alpha\nu}\;.\label{insul1}
\end{equation}
However, in the pure bosonic case, the conservation law \eqref{barg} is also valid and, because $T_f^{\mu\nu}=0$, is also a gauge invariant equation. Thus, the vanishing of the \emph{rhs} of \eqref{insul1} is not a requirement but a physical necessity. Then, there are three possible situations: i.) The media is linear. Then, the \emph{rhs} of \eqref{insul1} is proportional to $j_f^\mu$ (see \cite{Sobreiro:2011ie}) which vanishes by hypothesis. Then, \eqref{cond2} is trivially satisfied. ii.) The relation \eqref{cond2} is not automatically satisfied. In this case, the fact that the media is not isotropic/homogeneous is so strong that its constitutive relations cannot be described by $\chi$ as a Lorentz tensor. The material is so exotic that it induces a Lorentz breaking. This is clear from the fact that the current $T_b^{\mu\nu}$ describes a symmetry of the little group, \emph{i.e.}, a subgroup of the Lorentz group. iii.) The media is not that trivial that \eqref{cond2} is automatically satisfied. This condition has to be imposed as a subsidiary condition. 

We focus on the third situation (c). In that case we must take $\rho=\vec{j}_f=0$ at expression \eqref{h}. Then, we find\footnote{The first of \eqref{cond4} is then automatically satisfied.}
\begin{equation}
\vec{H}=\frac{1}{(\vec{\nabla}\cdot\vec{E})}\vec{D}\times\left(\frac{\partial\vec{E}}{\partial t}-c^2\vec{\nabla}\times\vec{B}\right)\;,\label{h2}
\end{equation}
It is clear that, this condition is satisfied if the media is linear (case (a)).

Another interesting situation occur if $\vec{\nabla}\cdot\vec{E}=0$. In that case equations \eqref{cond4} reduces to
\begin{eqnarray}
\vec{H}\cdot\left(\frac{1}{c^2}\frac{\partial\vec{E}}{\partial t}-\vec{\nabla}\times\vec{B}\right)&=&0\;,\nonumber\\
\vec{D}\times\left(\frac{1}{c^2}\frac{\partial\vec{E}}{\partial t}-\vec{\nabla}\times\vec{B}\right)&=&0\;,\label{cond5}
\end{eqnarray}
Thus, $\vec{H}\bot\vec{K}$ and $\vec{D}\parallel\vec{K}$, where $\vec{K}=\frac{\frac{1}{c^2}\partial\vec{E}}{\partial t}-\vec{\nabla}\times\vec{B}$. Then, $\vec{H}\bot\vec{D}$.

\section{Discussion}

In this letter we have studied the properties of spin currents in general material media. From first principles (Lorentz symmetry and the gauge principle) we were able to show that the non-conservation of spin-currents are intrinsically inherent to electrodynamics.

It was shown that the relativistic generalization of spin currents always obeys the non conservation law \eqref{div1}. This relation is obtained from first principles of gauge theories, in particular, the Lorentz symmetry and the gauge principle. Although we have not considered a direct interaction term between electrons and the media, this task is not difficult to be discussed at phenomenological level. In fact, one can add an extra term to \eqref{div1} that simulates the spin loss of electrons from their interaction with the media, exclusively. For instance,
\begin{equation}
\partial_\nu\mathcal{T}_f^{\nu\mu}=-\frac{e}{2}S_\nu
F^{\nu\mu}+\zeta C^\mu\;,\label{div1x}
\end{equation}
where $\zeta$ characterizes the strength of the electron-media coupling and $C^\mu$ is a four-vector that depends on the medium properties. Both quantities depend on the material properties and should be characterized experimentally. The major challenge is to obtain this extra term from first principles. However, if the extra term is determined, it should be possible to adjust the electromagnetic fields to compensate the loss, namely, $C^\mu=\frac{e}{2\zeta}S_\nu
F^{\nu\mu}$, producing a conserved spin-current, eventually.

Another interesting result is obtaining by applying the very same first principles to the electromagnetic fields. The result is the bosonic analogue of the electronic spin current \eqref{bw2} whose non-conservation is described by \eqref{div2}. In contrast with the fermionic case, the bosonic non-conservation law changes with respect to the vacuum case \cite{Sobreiro:2011ie}, specially because of the interaction between the electromagnetic field and the medium through the constitutive tensor. The consequence is that, for perfect insulators, a condition that depends on the medium properties and the fields are obtained, \eqref{h2}. Thus, no phenomenological term is needed.

A problem that emerges at the bosonic current analysys concerns its interpretation. At the vacuum \cite{Sobreiro:2011ie}, the vector density $\mathcal{N}^i$ is identically zero while the current $\mathcal{N}^{ij}\propto \vec{E}\cdot\vec{B}\delta^{ij}$. Thus, the current is a kind of measure of non-orthogonality between electric and magnetic fields which may define a flowing current, although its corresponding density vanishes identically. Consequently, there will be flow only if the current is not conserved, see the \emph{rhs} of the second of equations \eqref{spin2}. In that case, the flow will depend on the presence of free charges $\rho$ and currents $\vec{j}$. The general case is obviously richer because we can demand conservation through \eqref{cond4} and still consider nontrivial flows and densities. In fact, the definitions $\mathcal{N}^{i}\propto\left(\vec{D}\times\vec{E}\right)^i$ and $\mathcal{N}^{ij}\propto c^2D^iB^j+E^iH^j-E^kH^k\delta^{ij}$ are immediately interpreted as a measure of how anisotropic is the medium, otherwise the previous case is recovered. The inevitable conclusion is that, if one wishes to transport information through $\mathcal{N}^{ij}$, it is necessary to consider anisotropic media. In that case, conditions \eqref{cond4} can be used to manipulate the fields and the medium properties in order to produce a conserved current.

Finally, it is worth mention that the introduction of the constitutive tensor $\chi^{\mu\nu\alpha\beta}$ is a useful technique to transfer the exotic properties of the medium to a Lorentz tensor. However, if condition \eqref{h2} is not automatically satisfied, then the medium is so exotic that its properties cannot be acomodated by $\chi^{\mu\nu\alpha\beta}$. A Lorentz breaking is inevitable.

The natural continuation of the present analysys, which is beyond the scope of this work, is to apply the conservation conditions here obtained to specific systems, explore reliable experimental situations and pursue a deep understanding of the bosonic current.

\section*{Acknowledgements}

RFS is thankful to the Conselho Nacional de Desenvolvimento Cient\'{\i}fico e Tecnol\'ogico\footnote{RFS is a level PQ-2 researcher under the program Produtividade em Pesquisa, 304924/2009-1.} (CNPq-Brazil). The authors acknowledge Rodolfo Casana and Diego Gonz\'alez for fruitful discussions.

\end{document}